\documentstyle[epsfig,twocolumn,prd,aps]{revtex}

\newcommand{\K}{{\cal K}} \newcommand{\D}{{\cal D}}
\newcommand{\hBox}{{{\smash\Box}\hspace{-0.65em}\text{\raisebox{0.2em}{\^~}}\!}}

\begin{document} \draft

%%% Front matter

\wideabs{%%% Remove before submission!
\title{Gravitational Waves from Braneworld Inflation}
\author{Andrei Frolov and Lev Kofman}
\address{
  CITA, University of Toronto\\
  Toronto, Ontario, Canada, M5S 3H8\\
  {\rm E-mail: \texttt{frolov,kofman@cita.utoronto.ca}}
}
\date{September 2002}
\maketitle

\begin{abstract}
We investigate the generation of primordial gravitational waves from
inflation in braneworld cosmologies with extra dimensions. Advantage of
using primordial gravitational waves to probe extra dimensions is that
their theory depends only on the geometry, not on the microscopic
models of inflation and stabilization. $D(D-3)/2$ degrees of freedom of
the free bulk gravitons are projected onto the 3d brane as tensor,
vector and scalar modes. We found the following no-go results for a
generic geometry of a five (or D) dimensional warped metric with four
dimensional de Sitter (inflationary) slices and two (or one) edge of
the world branes: Massive KK graviton modes are not generated from
inflation (with the Hubble parameter H) due to the gap in the KK
spectrum; the universal lower bound on the gap is $\sqrt{3/2}\, H$.
Massless scalar and vector projections of the bulk gravitons are
absent, unlike in geometries with KK compactification.

A massless 4d tensor mode is generated from inflation with the
amplitude $H/M_P$, where $M_P$ is the effective Planck mass during
inflation, derived from the $D$ dimensional fundamental mass $M_S$ and
the volume of the inner dimensions. However, $M_P$ for a curved dS
braneworld may differ from that of the flat brane at low energies,
either due to the $H$-dependence of the inner space volume or
variations in the brane separation before stabilization. Thus the
amplitude of gravitational waves from inflation in braneworld cosmology
may be different from that predicted by inflation in 4d theory.
\end{abstract}

\pacs{PACS numbers: 04.50.+h, 98.80.Cq \hfill CITA-2002-22}
}
\narrowtext

%%% Main body

\section{Introduction} \label{sec:introduction}

Higher dimensional formulation of superstring/M theory, supergravity
and phenomenological models of the mass hierarchy has most obvious
relevance to cosmology. Early universe inflation can universally
amplify vacuum fluctuations of all light (minimally coupled to gravity)
degrees of freedom. Thus inflation serves as a powerful tool to probe
the fields associated with the extra dimensions.

In this paper we consider a class of models with extra dimensions
related to the braneworld scenarios. One of the most interesting recent
developments of the high energy physics phenomenology is the picture of
the brane worlds, where our 3+1 dimensional spacetime is the 3d brane
embedded in the higher dimensional space. In application to the very
early universe this leads to the brane world cosmology, where our 3+1
dimensional universe is the 3d curved brane embedded in the bulk. A lot
of efforts have been devoted to the search for the new features of the
background cosmological expansion, inflation in the brane world
settings, cosmological fluctuations which could probe the brane world
scenarios and others. Many of these problem are rather complicated. For
instance, the problem of gravitational waves which might be observed on
the brane requires treatment of all ${1 \over 2}D(D-3)$ degrees of
freedom of the bulk gravitons, the choice of their initial state, the
issue of the KK modes, transplanckian problem and others.

Cosmological gravitational waves are a very useful tool to probe the
very early universe, because their generation depends on the background
geometry but not on the details of the particle physics built in the
model. A historic example is the theory of cosmological gravitational
waves generated from the early de Sitter stage \cite{starob} which,
ironically, was suggested even before the microscopic models of
inflation were proposed.

Gravitons almost freely propagate through the expanding universe, and
may leave their prints on the power spectrum and polarization of the
cosmic microwave background. With the ever increasing accuracy of the
CMB anisotropy experiments, it is worth to enlarge the pool of the
primordial gravitational waves theories.

In the familiar model of inflation in the four dimensions,
gravitational waves are generated from vacuum fluctuations during
inflation. Gravitons have two polarizations, and temporal part of their
wave function $h_k(t)$ obeys an equation which is similar to the one
for the massless scalar field minimally coupled to gravity.

Late time asymptotics of $h_k(t)$, which is relevant to present
cosmological observations, contains long wavelength modes $k$. They
have (almost) scale free spectrum $k^3 \vert h_k\vert^2 \approx
\text{const}$. The dimensionless almost constant amplitude
\begin{equation} \label{ampl1}
  k^{3/2} h_k = {H \over M_P},
\end{equation}
is defined by the ratio of the only two scales in the problem, which
are related to the gravitational waves: the Hubble parameter during
inflation $H$ and the Planck mass $M_P$ at low energies. Usually the
Hubble parameter slightly varies during inflation, $H=H(t)$, which
brings slight tilt to the spectrum of gravitational waves (in the {\it
r.h.s.} of (\ref{ampl1}) $H \to H(k)$). In this paper we will assume
$H=\text{const}$, but will devote some discussion to the case when $H$
slightly varies.

In this paper we will study the problem of the primordial gravitational
waves generation from the early universe in the theories with large
extra dimensions. As opposed to excitations of some other fields,
gravitational waves propagate in the bulk. What information about the
bulk/brane geometry can be, in principle, observable on the 3d brane
through the gravitational excitations? The problem has the flavor of
the well known mathematical quest of ``how one can hear the shape of a
drum?'' \cite{drum}.

To make geometrical setting definite, we assume the brane world
inflation. We will consider geometry of five dimensional warped metric
with four dimensional de~Sitter slices
\begin{equation} \label{warp}
  ds^2 = dy^2 + A^2(y)\, ds_4^2 \ .
\end{equation}
The warp factor $A(y)$ is determined self-consistently by the five
dimensional Einstein equations, supplemented by the brane boundary
conditions. This formalism is briefly outlined in Section \ref{sec:2}.
For the sake of generality, we also present results for a more general
case of $D$ dimensional warped metric with $D-1$ dimensional de~Sitter
slices. The class of metrics (\ref{warp}) covers many important brane
world scenarios \cite{brane} including the Horawa-Witten theory, HW
with with curved branes, the Randall-Sundrum (RS) models, RSI model
with phenomenological stabilization of two branes, scalar sector of the
supergravity realization of the RS model and others. The presence of a
bulk scalar field $\phi$ (a dilaton) with the potential $V(\phi)$ or
several scalars is often assumed.

The warped metric (\ref{warp}) can be rewritten in
``conformally-factorized'' form
\begin{equation} \label{warp2}
  ds^2 = \Omega^{-2} \left( d\chi^2 + ds_4^2 \right),
\end{equation}
where conformal factor and conformal fifth coordinate are defined by
\begin{equation}
  \Omega(\chi) = \frac{1}{A(y(\chi))},\hspace{1em}
  \chi = \int \frac{dy}{A(y)}.
\end{equation}
In our treatment of gravitational waves, beginning in Section
\ref{sec:3}, we will extensively use conformal mapping of the whole
geometry (\ref{warp2}), bulk and branes on its edges, into a simpler
manifold with boundaries, with the metric
\begin{equation} \label{conf}
  d{\hat s}^2 = \Omega^2\, ds^2 = d\chi^2 + ds_4^2.
\end{equation}

There are several important papers on the formalism of metric
fluctuations \cite{form} and properties of the gravitational waves in
the brane world scenarios \cite{HHR,LMW,GKL}, and will try to make
connection with them in the course of the paper. We will work with the
metric in the form (\ref{warp2}), which allows separation of variables
in gravitational wave equation, as it will be shown in Section
\ref{sec:4}. Conformal mapping also significantly simplifies the issue
of the boundary conditions, as it will be explained in Section
\ref{sec:5}. Just as it was in the four dimensional case, in $D$
dimensional theory the graviton eigenfunctions $h_{k,m}(\chi)$ obey an
equation for the massless scalar minimally coupled to gravity.

In Section \ref{sec:6}, we show that the spectrum of nonzero KK modes
has a mass gap with the universal lower bound $ \sqrt{\frac{3}{2}}\,
H$, independently on the bulk bulk scalar fields potential $V(\phi)$
and the presence or absence of the second brane. This result
generalizes the finding of the ${3 \over 2}\, H$ gap in the
Randall-Sundrum II model with a single brane in the AdS bulk
\cite{LMW}.

In Section \ref{sec:graviton}, we show that the homogeneous zero KK
mode $m=0$ is degenerate, and can be described by only two physical
polarization degrees of freedom, while other three can be gauged by
zero by the residual coordinate transformation. Thus, the issue of the
gravi-scalar and gravi-vector modes in brane cosmology is resolved:
they do not contribute.

All together this allows us to conclude that the late time asymptotic
of the long-wavelength gravitational waves generated from the brane
inflation is defined by the two tensor polarization of the homogeneous
KK mode $m=0$, similar to what takes place in a case of inflation in
four dimensions. However, as we will discuss in Section \ref{sec:8}, it
does not mean that gravitational waves from the brane inflation in all
respects are identical to those in the four dimensional theory.
Essential difference arises in their amplitudes. Indeed, in five
dimensional theories there is additional scale $\ell$, associated with
the fifth dimension (brane separation in the HW theory, the AdS radius
in the RS models, etc.) With three scales in the problem, $\ell$, $H$
and $M_P$, one may have a dimensionless amplitude different from
(\ref{ampl1}). Indeed, for the braneworld inflation, one finds $k^{3/2}
h_k=\frac{H}{M_P} F(H\ell)$, with some function $F$. Remarkably, the
amplitude can be expressed economically in terms of the effective
Planck mass for curved (de Sitter) braneworld
\begin{equation} \label{ampl2}
  k^{3/2} h_k = {H \over M_{P,\text{eff}}} \ .
\end{equation}
Effective Planck mass $M_{P,\text{eff}}$ during inflation in the
braneworld scenario may differ from the current value $M_P \sim
10^{19}\,$GeV, therefore prediction for the gravitational wave
amplitude from brane inflation may differ from that from inflation in
4d theory.

In Section \ref{sec:9}, we will discuss implications of our results
to the observational test of the ratio of tensor to scalar modes,
transplanckian problem in the context of models with extra dimensions,
and other related issues.

We evoke the ``conformally-factorized'' frame (\ref{warp2}) mainly for
treatment of the gravitational waves. However, conformal mapping of the
brane worlds has merits of its own. In particular, we show that the
warp geometry with branes at the edges is conformally equivalent to the
portion of the spacetime with regular end of the world edges. For the
case of flat 4d slices of (\ref{warp2}), the world edges after
conformal mapping are empty.

\section{Background and gravitational perturbations}\label{sec:2}

In this section we outline the general formalism for treating
background geometry with fields (such as dilaton) in the bulk and on
the branes, and free perturbative gravitational waves in this
background. In general, there may also be scalar and vector bulk
perturbations as well as the perturbation in the brane position. We
will not consider them. However, we will address the question if bulk
gravitational perturbations generate scalar and vector perturbations at
the brane.

The brane worlds scenario with Einstein gravity and scalar fields is
described by the action
\begin{eqnarray}\label{action}
   S &=& \frac{1}{16\pi \kappa_D^2} \int \sqrt{-g}\, d^D x \,
          \left(R + 2{\cal L}_{\text{bulk}}\right)   - \\
     &-& \frac{1}{16\pi \kappa_D^2} \sum_a \int \sqrt{-h}\, d^{D-1} x \,
          \left(2 [\K] +2{\cal L}_{\text{brane}} \right) \ , \nonumber
\end{eqnarray}
where the first term describes the bulk and the second term is related
to the brane(s). The extrinsic curvature of the brane $\K_{\mu\nu}$ is
defined through the normal unit vector $n_A$ and the tangent vierbine
$e_{(a)}^B$
\begin{equation}\label{extr}
  \K_{\mu\nu} = e_{(\mu)}^A e_{(\nu)}^B \nabla_A n_B,
\end{equation}
and $\K$ is its trace. We use ``mostly positive'' signature and
curvature conventions of Misner, Thorne and Wheeler, and denote bulk
indices by capital Latin letters $A, B, ...$, brane indices by small
Greek letters $\mu, \nu, ...$, and brane hypersurfaces by $\Sigma_a$,
$a=1,2$. We use square brackets (e.g.\ $[\K]={\K}^+-{\K}^-$) to denote
jump of the geometrical quantity across the brane.

Einstein equations in the bulk are
\begin{equation}\label{bulk}
  R_{AB} - {1 \over 2}R g_{AB} = T_{AB},
\end{equation}
where the stress-energy tensor of the matter fields $T_{AB}$ is
described by the bulk Lagrangian ${\cal L}_{\text{bulk}}$. In
particular case of the minimally coupled scalar field with potential,
we take ${\cal L}_{\text{bulk}}=-{1 \over 2}(\nabla\phi)^2 - V(\phi)$ and
${\cal L}_{\text{brane}}=U(\phi)$, and the bulk stress-energy tensor is
\begin{equation} \textstyle
  T_{AB} = \phi_{,A} \phi_{,B} -
     \left(\frac{1}{2} (\nabla\phi)^2 + V(\phi)\right) g_{AB}.
\end{equation}
For 5d warped geometry (\ref{warp}) with scalar field the bulk Einstein
equations (\ref{bulk}) can be written as
\begin{eqnarray}\label{ein1}
\partial_y (A^{-1} \partial_y A)&=&-\frac{1}{4}(\partial_y \phi)^2-
\frac{1}{6} V(\phi)-(A^{-1} \partial_y A)^2 \ , \\ \label{ein2}
6 (A^{-1}\partial_y A)^2 &=&\frac{1}{2}(\partial_y \phi)^2-V(\phi)+6A^{-2}H^2 \ ,
\end{eqnarray}
where the second equation is a constraint. Comprehensive study of warp
geometry with a scalar field based on the qualitative analysis of the
dynamical system (\ref{ein1})-(\ref{ein2}) was given in \cite{FFK}.

The jump of extrinsic curvature across the brane surface is related to
the stress-energy tensor $S_{\mu\nu}$ of the matter contained on the
brane by Israel's junction conditions, which in the case of scalar
field with brane potential $U(\phi)$ are
\begin{equation}\label{junc}
  [\K_{\mu\nu} - \K g_{\mu\nu}] = U(\phi) g_{ \mu\nu}, \hspace{1em}
  [n \cdot \nabla\phi] = \frac{\partial U}{\partial\phi}.
\end{equation}
From here, the assumption of $Z_2$ symmetry around brane $\Sigma$ and
the warp geometry (\ref{warp}) implies the boundary condition for the
warp factor and the scalar field
\begin{equation}\label{bound}
  \left.\frac{\partial_y A}{A}\right|_{\Sigma} = -\frac{U}{2(D-2)}, \hspace{1em}
  \partial_y \phi|_{\Sigma} = {1 \over 2}\, \frac{\partial U}{\partial \phi}.
\end{equation}
In case of the RS models we have $U(\phi)=\pm \lambda$, where the sign
is different for the positive or negative tension $\lambda=\text{const}$
on the brane. The bulk potential is just 5d negative cosmological
constant $V(\phi)=-\Lambda$. In case of the HW model, $U(\phi)=e^{-\phi}$,
$V(\phi)=e^{-2\phi}$. Our calculations are valid for arbitrary $U(\phi)$
and $V(\phi)$. As we will see, the results of the paper can be
generalized for a more general case of several bulk scalars.

Free gravitational waves obey the linearized Einstein equations
\begin{equation} \label{eq:grav:einstein}
  \delta R^A_B=0.
\end{equation}
Linearized gravitational waves are described by the metric
perturbations $ h_{AB} \equiv \delta g_{AB}$, and the gravitational
wave equation is
\begin{equation} \label{gravwave}
  \Box h_{AB} - 2 \nabla^C \nabla_{(A} h_{B)C} + \nabla_A \nabla_B h + 2 R^C_{~(A} h_{B)C} = 0,
\end{equation}
where $h$ is a trace $h = {g}^{AB} {h}_{AB}$. In D dimensions,
gravitational waves have ${1 \over 2} D (D-3)$ independent physical
degrees of freedom. In discussions of brane world perturbations,
gravitational waves are often described by two components of the tensor
modes, associated with the brane isometry group. However, they
correspond only to the part of the bulk gravitational waves which have
5 degrees of freedom. In the formalisms of \cite{LMW,GKL} based on the
brane isometry group, one has to consider separately the pieces from
scalar and vector modes to complete the description of the 5d
gravitational waves.

In this paper we will deal with all five bulk graviton polarizations.
We will use the ``conformally-factorized'' form of the background
metric (\ref{warp2}), and impose the usual transverse-traceless gauge
for gravitational waves
\begin{equation} \label{eq:gauge:tt}
  \nabla^A h_{AB} = 0, \hspace{1em} h = 0.
\end{equation}
For this metric, it is convenient to fix some of coordinate freedom
assuming the following orthogonality condition
\begin{equation} \label{eq:gauge:orth}
  \nabla^A \Omega \, h_{AB} = 0.
\end{equation}
Similar coordinate gauge has been used in 4d cosmology where
$\Omega=\Omega(t)$ (the synchronous gauge). While the two conditions
(\ref{eq:gauge:tt}) and (\ref{eq:gauge:orth}) are not necessary
compatible for a most general background, they are compatible for the
brane-world metrics (\ref{warp}), since the orthogonality requirement
is with respect to a vector $u^A=\partial/\partial\chi$ that is
covariantly constant in the conformal frame $\hat\nabla_A u^A = 0$.

In the braneworld scenarios it is convenient to use the TT gauge
(\ref{eq:gauge:tt}), but one also can work with gravitational waves in
other gauges. It is interesting to follow the bulk gravitational waves
in the gauge which is usually used for the Kaluza-Klein
compactification (however, without $S^1$ symmetry in the compact $y$
direction). In the KK gauge, bulk gravitons are manifestly projected
into four-dimensional world as four dimensional TT tensor mode plus
scalar and vector modes. Consider the massless (homogeneous with
respect to $y$) free scalar and vector four dimensional components.
After substituting perturbations into the junction conditions
(\ref{junc}) we derive that all light scalar and vector projection of
the free bulk gravitational waves vanishes. This result will be
confirmed further by calculations in the gauge (\ref{eq:gauge:tt})
which we adopt in the rest of the paper.

\section{Gravitational Waves and Conformal Mapping}\label{sec:3}

There is a useful analogy between gravitation waves in the the braneworld
metric (\ref{warp2}), and those in the 4d cosmology with the metric
\begin{equation}\label{cosm}
  ds^2=a^2(\tau)\left(-d\tau^2+ds_3^2\right),
\end{equation}
where $ds_3=\gamma_{ij}dx^idx^j$ is the 3d metrics of constant
curvature $K=0,\pm1$, $a(\tau)$ is the scalar factor of the universe
and $\tau$ is a conformal time. It is well known \cite{KS84} that the
tensor perturbations in this metrics can be described by $h_{ij}$ which
can be conformally transformed and decomposed as $h_{ij}={1 \over a}
h_k(\tau)Q_{ij}$, where $Q_{ij}$ are two-component transverse-traceless
eigenmodes of $ds_3^2$, $(\nabla^2 + k^2)Q_{ij}=0$. Temporal part of
the eigenmodes obeys an equation similar to that of the minimally
coupled (nonconformal) scalar field
\begin{equation}\label{scal}
  \ddot h_k + \left(k^2 -{\ddot a \over a} +2K \right) h_k = 0,
\end{equation}
where dot is derivative with respect to $\tau$.

The warped metric (\ref{warp2}) is a higher dimensional version of
(\ref{cosm}) with time and one of the spatial coordinates exchanged.
Therefore we expect that gravitational wave equation (\ref{gravwave})
can be solved in the form $h_{\mu\nu} \propto h_m( \chi)Q_{\mu\nu}$,
where five gravitational polarizations are described by the
transverse-traceless eigenmodes $Q_{\mu\nu}$ of 4d de Sitter spacetime,
and $h_m(\chi)$ obeys an equation for minimally coupled bulk scalar
depending on conformal fifth coordinate $\chi$. Additionally, instead
of the Cauchy problem for graviton wavefunction $h_k(\tau)$ in the
metric (\ref{cosm}) we have Neumann boundary problem for graviton
wavefunction $h_m(\chi)$ in the metric (\ref{warp2}).

Let us begin with conformal transformation. All geometrical quantities
and operators evaluated with respect to the original metric $g_{AB}$
will be hatless, while those evaluated with respect to the conformally
transformed metric $\hat g_{AB} =\Omega^{2} g_{AB}$ will be denoted by
hat. Covariant derivative in physical frame will be $\nabla$, while
covariant derivative with respect to $\hat g_{AB}$ will be $\hat
\nabla$. Under general conformal mapping, background Ricci tensor
transforms as
\begin{eqnarray} \label{eq:conf:Ricci}
  &&
    R_{AB} = \hat{R}_{AB}
      + (D-2) \Omega^{-1} {\hat \nabla}_A {\hat \nabla}_B \Omega
  + \nonumber\\&&\hspace{3em}
      + \Omega^{-1} (\hBox\Omega) \hat{g}_{AB}
      - (D-1) \Omega^{-2} ({\hat \nabla}\Omega)^2 \hat{g}_{AB},
\end{eqnarray}
while its trace transforms as
\begin{equation} \label{eq:conf:R}
  R = \Omega^2 \hat{R} + 2(D-1) \Omega (\hBox\Omega) - D(D-1) ({\hat \nabla}\Omega)^2.
\end{equation}
For gravitational perturbations we have
\begin{equation} \label{grconf}
  \hat{h}_{AB} = \Omega^{2} h_{AB}.
\end{equation}
It can be shown that the gauge choice (\ref{eq:gauge:tt}) and
(\ref{eq:gauge:orth}) is conformally invariant. From the wave equation
in the physical frame (\ref{gravwave}), we obtain the gravitational
wave equation in the conformal frame
\begin{eqnarray} \label{eq:grav:conf}
  &&
    \hBox \hat{h}_{AB}
    - 2 \hat{\nabla}^C \hat{\nabla}_{(A} \hat{h}_{B)C}
    + 2 \hat{R}^C_{~(A} \hat{h}_{B)C}
  \nonumber\\&& \hspace{4em}
    - (D-2) \Omega^{-1} \Omega^{,C} \hat{\nabla}_C \hat{h}_{AB} = 0,
\end{eqnarray}
where no assumptions about the background spacetime were made yet
except the gauge (\ref{eq:gauge:tt}) and (\ref{eq:gauge:orth}).
Although equation in the conformal frame appears to have additional
term compared to (\ref{gravwave}), actually considerable simplification
occurs due to the simpler form of the background metric in the
conformal frame.

So far conformal transformation of the metric has been kept rather
general. Now we apply the result to the conformal mapping (\ref{conf})
with $\Omega=\Omega(\chi)$. The curvature of the metric $d\hat{s}^2$ is
determined by the curvature of the de Sitter portion, $\hat{R}_{\mu\nu}
= (D-2) H^2\, \hat{g}_{\mu\nu}$. Then equation (\ref{eq:grav:conf})
simplifies to
\begin{eqnarray} \label{eq:xd:grav}
  &&
    \hBox \hat{h}_{AB} - 2 \hat{\nabla}^C \hat{\nabla}_{(A} \hat{h}_{B)C} + 2 (D-2) H^2 \hat{h}_{AB}
  \nonumber\\&& \hspace{2.54em}
    - (D-2) \Omega^{-1}\Omega' \hat{h}_{AB}' = 0,
\end{eqnarray}
where prime denotes the derivative with respect to the extra dimension
$\chi$.

\section{Separation of variables in the gravitational waves equation}\label{sec:4}

In this section, we continue the analysis of the gravitational wave
equation (\ref{eq:xd:grav}). Since from the gauge (\ref{eq:gauge:orth})
it follows that $\hat h_{A\chi}=0$, only $\hat h_{\mu\nu}$ components
need to be considered. As $u^\mu=\partial/\partial\chi$ is a Killing
vector for the metric in the conformal frame, the dependence on the
extra dimension separates, and the gravitational perturbation can be
decomposed as
\begin{equation} \label{eq:xd:sep}
  \hat h_{\mu\nu} = \Omega^{{D-2} \over 2}h_m(\chi) Q^{(m)}_{\mu\nu}(x^{\rho}),
\end{equation}
where $m$ is an eigenvalue corresponding to the Kaluza-Klein modes. The
transverse-traceless conditions (\ref{eq:gauge:tt}) are inherited by
the $(D-1)$-dimensional tensor mode
\begin{equation} \label{eq:xd:gauge}
  \hat\nabla^{\mu} Q^{(m)}_{\mu\nu} = 0,\hspace{1em} Q^{(m)\mu}_{\mu} = 0,
\end{equation}
while no gauge conditions are placed on extra-dimensional mode $h_m$.

Upon substitution of the mode decomposition (\ref{eq:xd:sep}) into the
equation (\ref{eq:xd:grav}), the gravitational perturbation problem is
reduced to two decoupled differential equations. One of them is
equation for the $(D-1)$-dimensional tensor mode
\begin{equation} \label{eq:xd:H}
  \hBox Q^{(m)}_{\mu\nu} - 2 {\hat \nabla}^{\rho} {\hat \nabla}_{(\mu} Q^{(m)}_{\nu)\rho} +
  2(D-2)H^2 Q^{(m)}_{\mu\nu} = m^2 Q^{(m)}_{\mu\nu},
\end{equation}
and the other is an equation for the extra-dimensional scalar mode
\begin{equation} \label{eq:xd:schr}
  h_m''+ \left( m^2- V_{\text{eff}}\right) h_m = 0,
\end{equation}
where effective potential $V_{\text{eff}}$ is determined by the bulk geometry
\begin{equation} \label{eq:xd:veff}
  V_{\text{eff}} =
    - \frac{D-2}{2}\, \left( \frac{\Omega''}{\Omega}
    - \frac{D}{2}\, \frac{\Omega'^2}{\Omega^2} \right).
\end{equation}
There is some ambiguity in the choice of the separation constant $m^2$.
For instance, one can include the curvature of the 4d slices into
$m^2$, as it is usually done for the gravitational waves in open or
closed cosmology as in equation (\ref{scal}). Our choice of $m^2$ is
motivated by the following reason: it turns out, as we will soon see,
that $m^2$ in (\ref{eq:xd:H}) and (\ref{eq:xd:schr}) has the lowest
eigenvalue value $m^2=0$ and is always positive.

The equation (\ref{eq:xd:H}) is {\em exactly} the equation for linear
massive tensor perturbation of $(D-1)$-dimensional subspace. Thus, the
problem of gravitational waves in the braneworld metric (\ref{warp2})
is reduced to two decoupled equations, one for the massive tensor
polarizations $Q^{(m)}_{\mu\nu}$ on the $(D-1)$ spacetime of constant
curvature which we will analyze in detail in Section~\ref{sec:graviton},
and one for the minimally coupled scalar mode $h_m(\chi)$. These
equations are valid for a general case of an arbitrary warp factor
$\Omega(\chi)$, and for flat or curved branes of constant curvature.

The equation (\ref{eq:xd:schr}) has a convenient ``Schr\"odinger-type''
structure. The effective potential (\ref{eq:xd:veff}) can be represented
in different forms which are useful for different purposes. Using
conformal transformation of the Ricci scalar (\ref{eq:conf:R})
\begin{equation} \label{curvature}
{R \over \Omega^2}={\hat R}
 +2(D-1) \left( \frac{\Omega''}{\Omega}
    - \frac{D}{2}\, \frac{\Omega'^2}{\Omega^2} \right),
\end{equation}
the equation (\ref{eq:xd:schr}) can be written in the form
\begin{equation} \label{eq:xd:psi}
  h_m'' +\left[m^2 + \xi_c \left({R \over \Omega^2}-\hat R\right)\right]h_m=0,
\end{equation}
and we have an alternative expression for the effective potential
\begin{equation}\label{alt}
V_{\text{eff}}=\left(\frac{D-2}{2}\right)^2 H^2 - \xi_c\, \frac{R}{\Omega^2}.
\end{equation}
Numerical factor $\xi_c=\frac{1}{4} \frac{D-2}{D-1}$ corresponds to the
value of the coupling constant $\xi_c$ for a scalar field conformally
coupled with gravity. This is what one expects to obtain for the
equation of the minimally coupled (nonconformal) scalar field after
conformal transformation. Extra-dimensional graviton mode in the
physical frame corresponds to the minimally coupled massless scalar
field.

If we return to the warp factor $A(\chi)=1/\Omega(\chi)$ instead of
$\Omega(\chi)$ in (\ref{eq:xd:veff}), master equation (\ref{eq:xd:schr})
will be written as
\begin{equation}\label{form}
  h_m'' +\left[m^2 - {{D-2}\over 2} \left( \frac{A''}{A} +
 {{D-4}\over 2} \frac{A'^2}{A^2} \right) \right] h_m=0.
\end{equation}
This equation generalizes equation for the 4d cosmological
gravitational waves (\ref{scal}) (with the shifted choice of
the separation eigenvalue $m^2 \to k^2 +2K$).

Let us show that there is no tachyonic modes in the spectrum of KK
modes, i.e. $m^2$ is always positive. The equation (\ref{eq:xd:schr})
with the potential in the form (\ref{eq:xd:veff}) can also be written
in a factorized form
\begin{equation} \label{eq:xd:ss}
  -\D_-\D_+ h_m = m^2 h_m,
\end{equation}
where the differential operators are
\begin{equation} \label{eq:xd:D}
  \D_\pm = \partial_\chi \pm \frac{D-2}{2}\, (\ln\Omega)'.
\end{equation}
Simple manipulations with the equation (\ref{eq:xd:ss}) using
integration $\int d\chi$ between two branes, lead to the formula
\begin{equation}\label{eigen}
  m^2 = \frac{\int |\D_+ h_m|^2 d\chi}{\int |h_m|^2 d\chi} \ge 0,
\end{equation}
provided that boundary terms satisfy $(h_m^* \D_+ h_m)|_{\Sigma_a}=0$.
This condition is met for $Z_2$ symmetry.

\section{Boundary conditions and Conformal Mapping}\label{sec:5}

\begin{figure}
  \centerline{\epsfig{file=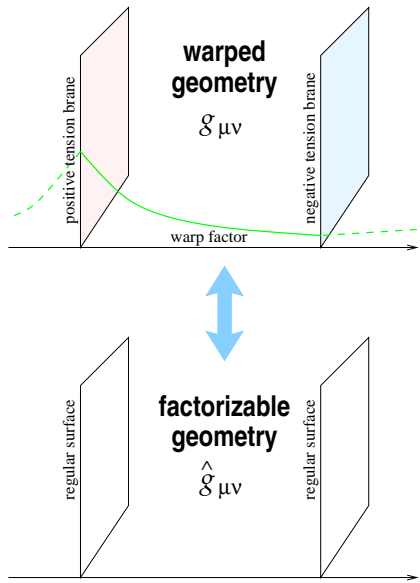, width=5.5cm}}
  \medskip
  \caption{
    Conformal transformation simplifies geometry. Upper panel
    represents warped geometry (\ref{warp}) with branes carrying
    tensions on the edges. Lower panel represents conformally-factorized
    geometry (\ref{conf}) with regular edges.
  }
  \label{fig:rs}
\end{figure}

The mass spectrum of the Kaluza-Klein modes is determined by a boundary
value problem in the extra dimension, and is obtained by imposing
desired boundary conditions on solutions of the equation (\ref{eq:xd:psi}).
In this section we discuss how the background and perturbative
quantities related to boundaries are mapped under conformal
transformation. The goal is to derive the boundary condition for the
mode function $h_m(\chi)$. If $Z_2$ reflection symmetry is assumed
across the brane, the constant tension brane is mapped to the edge of
the conformal space $\hat g_{AB}$. Remarkably, it turns out that after
conformal mapping the edge
{\it regular}, i.e. is not marked by the jumps in the curvature.
This conformal mapping is
sketched at Figure~\ref{fig:rs}.

The boundary condition for the gravitational wave $\hat h_{AB}$ in the
conformal frame at the empty $Z_2$-symmetric brane is simply $\hat
h_{\mu\nu}'=0$. Therefore, for the gravitational wave amplitude $h_m$,
we have a simple mirror-like (Neumann) boundary conditions
\begin{equation} \label{eq:xd:bc}
  \left.\left( \Omega^{{D-2} \over 2} h_m \right)'\right|_{\Sigma} = 0.
\end{equation}

In the rest of this section we shall demonstrate that conformal mapping
of the whole geometry, the conformally-factorized bulk metric
(\ref{warp2}) and its brane boundaries, brings it to the bulk metric
(\ref{conf}) with empty boundaries. This brane-world conformal mapping
property works for a broad class of warp geometries with the bulk
scalars with flat or de Sitter branes.

Let us consider the effect of the conformal transformation $d{\hat s}^2
= \Omega^2\, ds^2$ on the surface $\Sigma$ of co-dimension one embedded
in the spacetime manifold, and, in particular, on the extrinsic
curvature $\K_{\mu\nu}$. Suppose the surface to be parametrized as $x^A =
x^A(\xi^\mu)$, then the holonomic basis vectors tangent to the surface
$\Sigma$ have the same components in both physical and conformal frames
\begin{equation} \label{eq:conf:e}
  e_{(\mu)}^A = \frac{\partial x^A}{\partial \xi^\mu} = \hat{e}_{(\mu)}^A.
\end{equation}
Unit normal to the surface $\Sigma$ must be scaled, however, to
preserve the normalization
\begin{equation} \label{eq:conf:n}
  n^A = \Omega \hat{n}^A, \hspace{1em}
  n_A = \Omega^{-1} \hat{n}_A.
\end{equation}
The extrinsic curvature (\ref{extr}) of the surface $\Sigma$ under
conformal transformation behaves as
\begin{equation} \label{eq:conf:Kab}
  \K_{\mu\nu} = \Omega^{-1} \hat{\K}_{\mu\nu} - \Omega^{-2} (\hat{\nabla}\Omega \cdot \hat{n})\, \hat{g}_{\mu\nu},
\end{equation}
with its trace transforming as
\begin{equation} \label{eq:conf:K}
  \K = \Omega \hat{\K} - (D-1) (\hat{\nabla}\Omega \cdot \hat{n}).
\end{equation}

While tangential derivatives of the conformal factor must always be
continuous across the surface for induced metric to match, the normal
derivative need not be so. In particular, if the brane-world has the
$Z_2$ symmetry across the brane, $\K^-_{\mu\nu}=-\K^+_{\mu\nu}$ and
$\hat{\K}^-_{\mu\nu}=-\hat{\K}^+_{\mu\nu}$, and so the normal
derivative of the conformal factor flips sign $(\hat{\nabla}\Omega
\cdot \hat{n})^- = - (\hat{\nabla}\Omega \cdot \hat{n})^+$. In
geometrical terms, this means that the gradient of the conformal factor
$\hat{\nabla}\Omega$ experiences mirror reflection on the brane.

Now let us consider what happens to junction conditions (\ref{junc})
under conformal transformation. The left hand side transforms as
\begin{equation}\label{jump3}
[\K^{\mu}_{\nu}-\K\delta^{\mu}_{\nu}] = \Omega [\hat{\K}^{\mu}_{\nu}-\hat{\K}\delta^{\mu}_{\nu}]
+ (D-2)[\hat{\nabla}\Omega \cdot \hat{n}]\, \delta^{\mu}_{\nu}.
\end{equation}
In the physical frame we have
$[\K^{\mu}_{\nu}-\K\delta^{\mu}_{\nu}]=U(\phi)\delta^{\mu}_{\nu}$,
where $U(\phi)$ is the scalar field potential at the brane. On
the other hand, the last term in equation (\ref{jump3}) gives us
$(D-2)[\hat{\nabla}\Omega \cdot \hat{n}] = 2(D-2)\partial_\chi\Omega|_{\Sigma}=
-2(D-2)\frac{\partial_y A}{A}|_{\Sigma}$, where $A$ is the warp
factor. As can be seen from the junction condition in the form
(\ref{bound}), these two terms in (\ref{jump3}) are identical, and
cancel. Therefore for the junction condition in the conformal frame we
simply have
\begin{equation}\label{jump4}
[\hat{\K}^{\mu}_{\nu}]=0.
\end{equation}
This means that in the conformal frame the brane becomes {\it a
regular surface}, which means no special junction conditions or
boundary terms are needed. The flat branes with $H=0$ in the conformal frame
becomes an empty regular surfaces.

\section{Spectrum of KK modes}\label{sec:6}

Equipped with the extra-dimensional mode equation (\ref{eq:xd:schr})
and its boundary conditions (\ref{eq:xd:bc}), we can analyze its
solution $h_m(\chi)$, in particular, the spectrum of KK modes for the
brane world warped geometries (\ref{warp2}).

First we notice the presence of zero mode $m^2=0$ homogeneous KK solution,
which immediately follows from the equation in the form (\ref{eq:xd:ss})
\begin{equation}\label{homog}
  h_m(\chi)=\Omega^{-\frac{D-2}{2}},\hspace{1em} m^2=0.
\end{equation}

\begin{figure}
  \begin{center}\begin{tabular}{cc}
    \epsfig{file=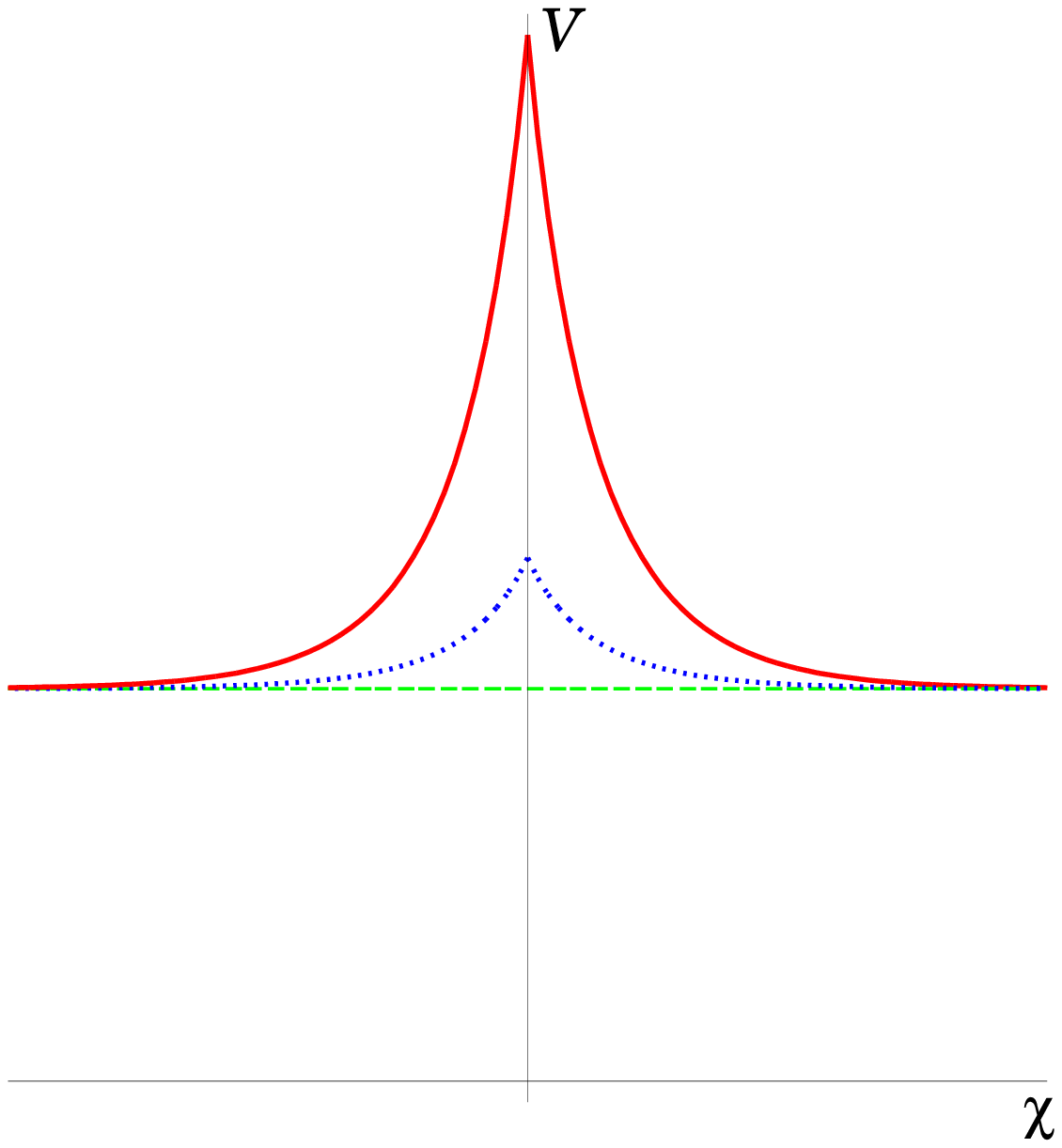, width=4cm} &
    \epsfig{file=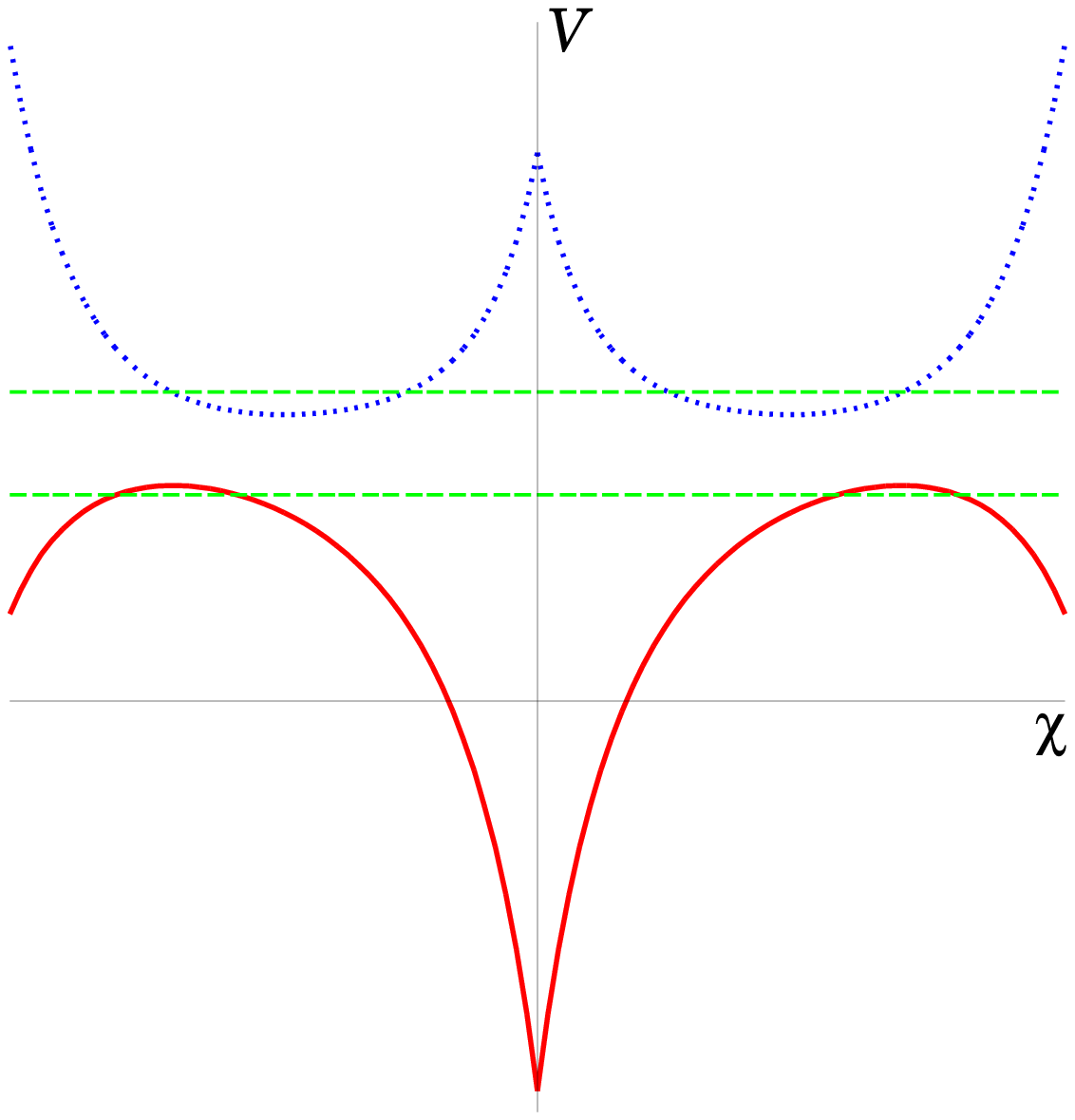, width=4cm} \\
  \end{tabular}\end{center}
  \caption{
    Schr\"odinger potentials $V_{\text{eff}}$ (red solid line) and
    $\tilde V_{\text{eff}}$ (blue dotted line) for the RS model (left)
    and HW model (right) in the periodic brane-worlds between two
    branes. Upper horizontal (green) line correspond to the gap ${3
    \over 2}H$ in the RS model; lower horizontal (green) line
    correspond to the generic gap $\sqrt{3 \over 2}\, H$.
  }
  \label{fig:pot}
\end{figure}

For non-zero KK modes we have to study the eigenvalues of the
Schr\"odinger-type equation (\ref{alt}), which depend on the the shape
of the effective potential $V_{\text{eff}}(\chi)$. Figure~\ref{fig:pot}
illustrates this function for two examples of the braneworld models.
Left panel corresponds to the Randall-Sundrum models (with one or two
branes) with negative cosmological constant $\Lambda$ in 5D AdS bulk.
Right panel shows $V_{\text{eff}}$ in the warp geometry model with bulk
scalar field potential $V(\phi)=V_0 e^{-2\phi}$, corresponding to the
HW model.

At first glance, the spectrum of massive KK modes crucially depends on
$V_{\text{eff}}(\chi)$, and consequently, on the braneworld theory.
However, it turns out that in general case one encounters the gap in
the spectrum of KK modes always greater or equal to $\sqrt{3 \over 2}\,
H$. Exact numerical factor in the front of $H$ can be slightly
different (but greater) in different theories.

First, let us inspect the equation (\ref{alt}). If the bulk curvature
is negative, $R \leq 0$, then the effective potential in the equation
for the KK amplitude $h_m(\chi)$ has a lower bound
\begin{equation}\label{gap}
  V_{\text{eff}} \geq \left(\frac{D-2}{2}\right)^2 H^2, \hspace{1em} R \leq 0,
\end{equation}
and the mass gap $m \ge \frac{D-2}{2}\, H$ appears in the spectrum. This
takes place in both RS models (with one or two branes) with the AdS
background supported by the negative bulk cosmological constant. The
presence of the gap ${3 \over 2} H$ in the RS model with a single brane
was noticed earlier in \cite{LMW} and in the domain wall in AdS model
in \cite{HHR}.

If the bulk curvature is positive, the potential $V_{\text{eff}}$ can
be smaller than $\left(\frac{D-2}{2}\right)^2 H^2$ or even become
negative. To analyze the mass spectrum in this, more general, case we
will use the following trick. For a given system of eigenfunctions
$h_m$, let us introduce adjoint system $\tilde h_m = \D_+ h_m$. While
$h_m$ obey the equation in the form (\ref{eq:xd:ss}), the eigenfunctions
$\tilde h_m$ obey the adjoint equation
\begin{equation} \label{eq:xd:sss}
  -\D_+\D_-{\tilde h_m} = m^2 \tilde h_m.
\end{equation}
In the terminology of SUSY quantum mechanics, the system $\tilde h_m$
is a superpartner of $h_m$. The essential feature of adjoint
representation for us is that the set of eigenvalues $m^2$ is the same
for both systems $h_m$ and $\tilde h_m$.

Now consider equation (\ref{eq:xd:sss}). This equation again can be
rewritten in the form of the Schr\"odinger equation
\begin{equation} \label{eq:xd:schrs}
  \tilde h_m''+ \left( m^2- \tilde V_{\text{eff}}\right)\tilde h_m = 0,
\end{equation}
where, using the Einstein equations (\ref{ein1},\ref{ein2}) for the
bulk scalar field, the effective potential $\tilde V_{\text{eff}}$ for
$\tilde h_m$ can be written in the form
\begin{equation} \label{alts}
  \tilde V_{\text{eff}} = \frac{D-2}{2}\, H^2
      + \frac{A^2}{2} (\partial_y \phi)^2
      + \frac{(D-4)(D-2)}{4} (\partial_y A)^2.
\end{equation}
In case of several bulk scalars $\phi_i$ gradient term $(\partial_y
\phi)^2$ in (\ref{alts}) is replaced by positive term $\sum_i
(\partial_y \phi_i)^2$. This ``supersymmetric'' effective potential
$\tilde V_{\text{eff}}$ for two examples, the RS and HW models, is
plotted in Figure~\ref{fig:pot} alongside with $ V_{\text{eff}}$. From
(\ref{alts}) one can immediately see that for an arbitrary scalar field
theory
\begin{equation}\label{gap1}
  \tilde V_{\text{eff}} \geq \frac{D-2}{2}\, H^2.
\end{equation}
Therefore, for $D=5$ the spectrum of KK modes has the mass gap of at
least $\sqrt{3 \over 2}\, H$.

The presence of the gap means that there will be no KK gravitational
modes (except $m=0$ mode) which will contribute to the late time
cosmological gravitational waves. This is because the allowed modes are
heavy, $m \geq \sqrt{\frac{3}{2}}\, H$ for $D=5$, relative to the
Hubble parameter during inflation. As we will argue in the Section
\ref{sec:graviton}, the heavy KK modes do not survive the late-time
long wavelength cosmologically relevant asymptotics.

\section{Four-dimensional massive tensor modes}\label{sec:graviton}

The gravitational waves from the brane-world inflation can be
factorized into the Kaluza-Klein modes $h_m(\chi)$ which obey the
equation (\ref{eq:xd:schr}), and $(D-1)$ dimensional tensor modes
$Q_{\mu\nu}^{(m)}(x^{\rho})$ which obey the equation (\ref{eq:xd:H}).
In this section we consider $(D-1)$ tensor modes. For cosmological
applications we will concentrate on $D=5$, i.e. on four-dimensional
tensor modes.

De~Sitter manifold can be covered by different coordinates. It is
customary in the inflationary cosmology to use a spatially flat dS
slicing
\begin{eqnarray} \label{DS1}
  ds_{\text{dS}}^2
    &=& - dt^2+ e^{2Ht}dx_{i}dx^{i} \\
    &=& a^2(\tau)\left( -d\tau^2+ dx_{i}dx^{i} \right), \nonumber
\end{eqnarray}
where $i=1,2,3$, $a(\tau)=-1/H\tau$. These coordinates are very
convenient to study scalar fluctuations or four-dimensional massless
gravitons. In paper \cite{LMW}, they were used to study massive
transverse traceless tensor modes of the isometry group $R^3$ of a
flat three dimensional slices of (\ref{DS1}). As we already noted, this
approach is not fully consistent, since these modes have only two
polarizations and do not describe the five component bulk gravitons.

To deal with the massive transverse traceless tensor modes of dS space
with the bigger isometry group $SO(4,1)$ we shall use the global de
Sitter map
\begin{eqnarray}\label{DS}
  ds_{\text{dS}}^2 &=& - dt^2+ {1 \over H^2} \cosh^2{(Ht)}\, d\Omega_3^2, \\
 &=& { 1 \over {\cos^2 \tau}}\left(-d \tau^2 + d\Omega_3^2 \right); \nonumber\\
d\Omega_3^2&=& d\psi^2 + \sin^2 \psi (d \theta^2+\sin^2 \theta d \phi^2). \nonumber
\end{eqnarray}
Equation (\ref{eq:xd:H}) for the metric (\ref{DS}) is the equation for
tensor eigenfunctions of the Laplace-Beltrami operator on de Sitter
spacetime. The iteration scheme to construct tensor eigenfunctions from
scalar hyper-spherical harmonics was developed in \cite{higuchi}.

First, we will consider the massless mode $m^2=0$. This is very
important special case corresponding to the homogeneous KK mode. In
this case the equation (\ref{eq:xd:H}) corresponds {\it exactly} to the
equation for the usual four dimensional massless tensor mode. Tensor
mode $Q_{\mu\nu}^0$ has not five but only two transverse traceless
degrees of freedom, corresponding to the two polarizations of
four-dimensional graviton. Technical reason for this degeneracy is that
there is a residual gauge freedom of the synchronous gauge in four
dimensions which can be used to zero three components of tensor
perturbations. When $m^2$ is non-zero, the degeneracy is lifted.

In $m^2=0$ case solution of the equation (\ref{eq:xd:H}) can be further
factorized and written as \cite{higuchi,allen}
\begin{eqnarray}\label{homo}
 Q_{0\nu}^0&=&0 \ , \,\,\,
  Q_{ij}^0={ N \over {a(\tau)}}
 f_{nlm}(\tau) \, T^{nlm}_{ij}(\psi, \theta, \phi) \ , \label{homo1} \\
% \nonumber\\
f_{nlm}(\tau)&=& { 1 \over{ \sqrt{2n(n+1)(n+2)}}} \nonumber\\
&\times&\left((n+2) \cos \tau -e^{-i\tau} \right)e^{-i(n+1)\tau} \ ,
\end{eqnarray}
where $N$ is normalization constant independent on $(n, l, m)$,
$T^{nlm}_{ij}(\psi, \theta, \phi)$ are transverse traceless tensor
harmonics of three-sphere $d\Omega_3^2$ explicitly derived in
\cite{higuchi}, which have two independent degrees of freedom.

The key point for our discussion is that in the late time limit $t \to
\infty$, $\tau \to \pi/2$, the amplitude $f_{nlm}(\tau)$ freezes out at
the constant value $f_{nlm}(\tau) \to { 1 \over{\sqrt{2n(n+1)(n+2)}}}$.
Similar treatment of massless gravitons in the planar coordinates
(\ref{DS}) gives us frozen amplitude $f_{\vec k}\to\frac{1}{(2k)^{3/2}}$
where $\vec k$ is 3D comoving momentum. Therefore KK zero modes are, in
principle, cosmologically observable, depending on their amplitude $N$.

Let us turn to the massive KK modes with $m^2 > 0$. In this case there
is no degeneracy of bulk graviton degrees of freedom, and simple
factorization (\ref{homo}) is not applicable. Massive tensor KK modes
in five dimensions are described by all five components. As we shown in
the previous section, in the brane inflation scenario massive gravitons
are heavy, with the mass value which exceeds the universal threshold
$\sqrt{3 \over 2}\, H$. The key point about heavy gravitons is that they
are not generated during inflation.

This is similar to the properties of massive scalar field fluctuations
during inflation. Indeed, familiar scalar eigenfunctions of the
Laplace-Beltrami operator in the de Sitter geometry in the planar
coordinates are given in terms of the Hankel functions
\begin{equation}\label{temp1}
  f^{(m)}_k(\tau)=\frac{\sqrt{\pi}}{2} H \vert\tau \vert^{3/2} {\cal H}^{(1)}_\lambda(k \tau) \ ,
\end{equation}
with the index
\begin{equation}\label{index}
  \lambda= \sqrt{ \frac{9}{4}-\frac{m^2}{H^2}} \ .
\end{equation}

The physical wavelength of the quantum fluctuations is stretched out
with time as $a(\tau) \over k$. Cosmologically interesting fluctuations
have very small comoving momenta $k \to 0$. The amplitude of interest
corresponds to the late-time asymptotics of conformal time $\tau \to 0$
(or $t \to \infty$). The late-time long-wavelengths asymptotic $k \tau
\to 0$ of $f^{(m)}_k(\tau)$ crucially depends on the value of parameter
${m^2 \over H^2}$. For massless or very light modes, ${m^2 \over
H^2}\ll 1$, the fluctuations are frozen out at the level
\begin{equation}\label{asymp}
  f^{(m)}_k(\tau)\simeq \frac{H}{\sqrt{2}k^{3/2}}
\end{equation}
as far as its physical wavelength exceeds the Hubble radius ${a(\tau_k)
\over k} > H^{-1}$. On the contrary, the heavy modes $m^2 \geq H^2$ do
not freeze out but oscillate with decreasing amplitude
\begin{equation}\label{asymp1}
f^{(m)}_k(\tau)\propto \tau^{i \sqrt{\frac{m^2 }{H^2}-\frac{9}{4}}+\frac{3}{2} }
\end{equation}
as $\tau \to 0$.

Therefore we conclude that {\it the heavy KK modes do not contribute to
the cosmologically interesting late-time long-wavelengths asymptotics
of the gravitational waves and scalar fluctuations}. This conclusion is
similar to the result for the scalar field fluctuations in the model
with inflation in outer space in theories with extra dimensions
(anisotropic inflation) with the KK compactification \cite{KS87}. There
it was shown that heavy KK modes do not contribute to cosmological
fluctuations generated from inflation.

\section{Braneworld Dimensional Reduction and Normalization of the Gravitational Wave}\label{sec:8}

Einstein gravitational action in $D$ dimensions is
\begin{equation}\label{full}
  S_D = M_S^{D-2} \int R[g_{AB}]\, \sqrt{-g}\, d^D x
\end{equation}
plus surface terms, where $M_S$ is $D$-dimensional fundamental mass. If
we can factorize the metric $g_{AB}$ into four dimensional outer space
and $(D-4)$ dimensional inner space of volume $V_{\text{inner}}$, then
the action (\ref{full}) will be reduced to the effective
four-dimensional Einstein action
\begin{equation}\label{four}
  S_4=M_{P,\text{eff}}^2 \int R[g_{\mu\nu}] \sqrt{-g}\, d^4x,
\end{equation}
where the effective four-dimensional Planck mass $M_{P,\text{eff}}$
is related to the fundamental mass $M_S$ via formula
\begin{equation}\label{planck}
  M_{P,\text{eff}}^2=M_S^{D-2} V_{\text{inner}}.
\end{equation}

Let us follow how this procedure works for particular example of
conformally-factorizable $4+1$ metric (\ref{warp}) of the braneworld
scenario. First we perform conformal mapping (\ref{eq:conf:R}) and
assume $\hat R$ does not depends on the fifth dimension $\chi$. By
doing this, we restrict ourselves to zero-mode (with respect to $\chi$)
gravitational perturbations, which is the subject of cosmological
interest. By integrating extra-dimensional dependence out, we have
\begin{equation}
  S_D = M_S^{D-2} \int \Omega^{2-D}(\chi) d\chi
        \int \hat{R}[\hat{g}_{AB}] \sqrt{-\hat{g}}\, d^{D-1} x + \dots,
\end{equation}
where we have omitted derivative of $\Omega$ terms for brevity.
The effective action for the four-dimensional theory is then
\begin{equation}
  S_4 = M_{P,\text{eff}}^2 \int R[g_{\mu\nu}] \sqrt{-g}\, d^4 x + \dots,
\end{equation}
where effective four-dimensional Planck mass is
\begin{equation}
  M_{P,\text{eff}}^2 = M_S^3 \ell_{\text{eff}}
\end{equation}
and $\ell_{\text{eff}}$ is the effective size (volume) of the inner
dimension. For braneworld scenario with compact inner dimension $\chi$,
it is
\begin{equation}\label{size}
  \ell_{\text{inn}} =
    \int\limits_{\chi_1}^{\chi_2} \Omega^{-3}(\chi) d\chi =
    \int\limits_{y_1}^{y_2} A^2(y) dy.
\end{equation}
The warp factor $A(y)$, in principle, depends on the brane curvature
(defined by $H$) through the Einstein equation (\ref{ein2}). In
particular, it is different for de Sitter brane geometry with the
Hubble parameter $H$ and for the flat brane with $H=0$. Independently,
physical separation between curved branes may not be the same as the
separation of flat branes at low energies. Indeed, in the brane-world
cosmology energy-momentum tensor at the brane is time dependent, and so is
the extrinsic curvature, i.e. the brane embedding. Therefore the
effective volume $\ell_{\text{eff}}$ of the extra dimension is changing
in the course of cosmological evolution between inflation and present
universe.

Thus, {\it the effective Planck mass $M_{P,\text{eff}}$ should be
expected to be different during inflation and in the present universe.}

Let us return to the gravitational waves. In the actions (\ref{full}),
(\ref{four}) we split the curvature into the background part plus
gravitational wave perturbations. To construct the graviton field
theory we keep perturbations up to the second order. Contribution
to the action from the second order graviton perturbations reads as
\begin{equation}\label{second}
  \delta S_5 = M_S^3 \int \Omega^{-3} \left[ \frac{1}{4}
    \hat{h}_{AB;C} \hat{h}^{AB;C} + \dots \right] \sqrt{-\hat g}\, d^5 x.
\end{equation}
As we found in the previous section, only zero KK mode contributes to
the interesting cosmological effects. For these modes integrant in
(\ref{second}) does not depend on the fifth dimension, and we obtain
\begin{equation}\label{second1}
  \delta S_4 = M_{P,\text{eff}}^2 \int \left[ \frac{1}{4}
    h_{\mu\nu;\rho}h^{\mu\nu;\rho} + \dots \right] \sqrt{-g}\, d^4 x.
\end{equation}
Thus, we reduce the problem back to the four-dimensional theory of
gravitational waves from inflation. The standard approach is to rewrite
action (\ref{second1}) in terms of scalar field $\varphi$ with the
canonical kinetic term, $\varphi= \frac{M_{P,\text{eff}}}{\sqrt{2}} h$,
and quantize the scalar field $\varphi$ in the de Sitter background.
Scalar field vacuum fluctuations are amplified by inflation to the
magnitude $k^{3/2}\varphi_k=\frac{H}{\sqrt{2}}$ (see also (\ref{asymp})). As
a result, we obtain that long-wavelength gravitational waves are
generated from inflation with the scale free spectrum and the amplitude
\begin{equation}\label{amplituda}
  k^{3/2} h_k=\frac{H}{M_{P,\text{eff}}}.
\end{equation}
Thus, the only difference with the results for the gravitational wave
amplitude in the usual four-dimensional theory is the effective
Planck mass $M_{P,\text{eff}}$ instead the fixed $M_P$, namely
\begin{equation}\label{ampl}
  k^{3/2} h_k=\frac{H}{M_P} \sqrt{ \frac{\ell_{\text{now}} } {\ell_{\text{inf}}}},
\end{equation}
where $\ell_{\text{now}}$, $\ell_{\text{inf}}$ are the effective sizes
of extra dimension at the inflationary stage and today.

Let us discuss how $M_{P,\text{eff}}$ at the stage of inflation in the
braneworld scenario may be different from the current $M_P$. It is
instructive to consider the simple Randall-Sundrum model with a single
inflating brane in five dimensional AdS bulk. There we have the
conformal factor $\Omega(\chi)= \frac{\sinh H\chi}{H\ell}$. Then
explicit integration in (\ref{size}) gives \cite{LMW}
\begin{equation}\label{func}
  \ell_{\text{eff}} =
    \ell \left\{
       \sqrt{1+H^2\ell^2} + H^2\ell^2 \ln
       \left( \frac{\sqrt{1+H^2\ell^2} - 1}{H\ell} \right)
    \right\}.
\end{equation}
For the ratio $\frac{\ell_{\text{now}}}{\ell_{\text{inf}}}$ involved in
formula (\ref{ampl}) we have the asymptotics
\begin{equation}
  \frac{\ell_{\text{now}}}{\ell_{\text{inf}}} \rightarrow
    \left\{
      \begin{array}{cl}
        1 & \text{for} H\ell \ll 1 \\
	\frac{3}{2} (H\ell) & \text{for } H\ell \gg 1
      \end{array}
    \right..
\end{equation}
If brane inflation occurred at scales smaller than the curvature of the
extra dimension (i.e.~$H\ell \gg 1$), and today universe is almost flat
(i.e.~$H\ell \ll 1$), the gravitational wave amplitude generated by
inflation is enhanced
\begin{equation}
 k^{3/2}h_k =
  \frac{H}{M_{P,\text{eff}}} =
  \frac{H}{M_P} \sqrt{\frac{3}{2} H\ell}
\end{equation}
compared to our expectation from standard cosmology.

Another example where $M_{P,\text{eff}}$ is changing in the braneworld
scenario was considered in \cite{GKL}. It can be shown that the model
of \cite{GKL} effectively corresponds to the case of two parallel
inflating branes separated by scale $\ell$ embedded in flat empty five
dimensional space. There it was assumed that $H\ell \ll 1$.

We are not aware of convincing arguments why $H\ell$ cannot be of order
of unity or larger. The case of $H\ell \ll 1$ corresponds to two
(almost) parallel branes separated by the scale $\sim \ell$ which is
small compared to their curvature radius $H^{-1}$. It looks natural to
have as initial conditions two branes parallel within the casually
connected region \cite{KL}. From point of view of four dimensional
observer this is the region $\sim H^{-1}$, However, casually connected
region in five dimensional space is defined not by $H$ but by five
dimensional curvature. For instance, in the RSI scenario with the AdS
bulk already for $H\ell\sim 1$ the function $\ell_{\text{eff}}$ in
(\ref{func}) drops by factor $2$ (it is a very steep function of
$H\ell$). The case $H\ell \geq 1$ gives us interesting possibility to
have significant departure from the standard prediction of the
inflation in the four dimensional theory.

So far we discussed the amplitude of the gravitational waves at the
stage of inflation. Will formula (\ref{amplituda}) be the final answer
for the gravitational wave amplitude in the late time cosmology? The
reason for this question is the following. The amplitude $h_k$ may
somehow be affected during cosmological evolution between inflation and
the latest stages because $M_{P,\text{eff}}$ may change its value
between inflation and current value (say, after stabilization after
inflation).\footnote{We thank Shinji Mukohyama for this comment.}
Although equation for free gravitational waves does not depends on the
gravitational coupling, situation with the varying $M_P$ is more
delicate. One may model this situation in framework of the Brans-Dicke
theory, where BD scalar has two constant but different asymptotics.
Simple estimations with this approach indicates that there is no
additional variation of the amplitude $h_k$. This issue deserves
further investigation.

\section{Discussion}\label{sec:9}

Among the many features of superstring/M theory, the existence of extra
dimensions has the most direct relevance to cosmology. Gravitons
propagate in all -- outer and inner -- dimensions. Bulk gravitons in
$D$ dimensions have $\frac{1}{2}D(D-3)$ components. Only two of them
appear as a transverse traceless tensor mode from the point of view of
a four dimensional observer, while the others are projected into the
$3+1$ dimensional world as (gravi)scalars and (gravi)vectors.

Vacuum fluctuations of all light (minimally coupled) degrees of freedom
are unstable during inflation and appear as long-wavelength
perturbations in late time cosmology. Thus, outer space inflation acts
like an amplifier and stretcher of light modes. Using inflation as a
tool, we may probe extra dimensions by examining light modes associated
with extra dimensions which may be produced by inflation.

It is especially appealing to investigate gravitational waves in the
context of extra dimensions plus inflation. This is because the theory
of small gravitational fluctuations depends on the underlying geometry
but not on the microscopic model of its realization. Additionally,
gravitons propagate almost freely throughout the evolution of the
universe, thus carrying information about the primordial universe.

There are two distinct types of scenarios with compactification of
extra dimensions from the point of view of the theory of primordial
gravitational waves. One type is Kaluza-Klein compactification. The
theory of gravitational waves from inflation in models with KK
compactification will be considered separately. Another type of model
arises from the braneworld scenarios with heavy branes (including a
single brane in AdS bulk or compact warped inner space in between two
branes). In this paper we investigated primordial gravitational waves
in generic geometries of the warped inner space with outer space
(brane) inflation.

We found the following no-go results for gravitational waves from
inflation in the generic case of these braneworld scenarios:

Massless mode of bulk gravitons is projected to the usual four
dimensional two component TT tensor mode, while their scalar and vector
projections are absent. This is different from KK compactification
where light bulk gravitons are projected to tensor, vector and scalar
massless components in the four dimensional world.

Massive KK modes are not generated from inflation due to the gap in the
KK spectrum. The gap is present in all models including models with two
branes. The universal lower bound of the gap is
\begin{equation}
  \Delta m \geq \sqrt{\frac{3}{2}}\, H.
\end{equation}
This result generalizes the finding of the gap (of magnitude
$\frac{3}{2} H$) for the RS model with a single brane embedded in AdS
\cite{LMW}.

Among all massless projections of the bulk gravitons, only the four
dimensional TT massless modes are generated from brane inflation. As in
the standard four dimensional theory, their spectrum is scale free, but
the amplitude is given by the formula
\begin{equation}
  k^{3/2} h_k=\frac{H}{M_{P,\text{inf}}},
\end{equation}
where $M_{P,\text{inf}}$ is the four dimensional Planck mass for the
curved (de Sitter) observable brane. Suppose the inner dimensions are
stabilized after inflation (say due to SUSY breaking). This is not too
radical an assumption bearing in mind that the usual energy scale of
inflation is $H \sim 10^{13}$ GeV and the usual energy scale of SUSY
breaking is $\sim 10^{11}$ GeV. Therefore theoretically it is not
necessary that $M_{P,\text{inf}}$ during inflation coincides with
present day low energy $ M_{P}$. The difference may come from variation
of the size of the inner dimension(s) (i.e. the limits of integration
in formula (\ref{size}), or from an imprint of the brane curvature $H$
on the warp factor $A(y)$. A change in $M_{P,\text{eff}}$ by a factor
of $2$ is already interesting. Thus in general, the connection between
gravitational wave amplitudes and energy scale of inflation $H$ in
theories with extra dimensions may be different from that in four
dimensional theory.

Most important test of the primordial cosmological fluctuations from
inflation is the ratio between the amplitude of tensor and scalar
perturbations $T/S$. To calculate scalar perturbations, one needs to
specify the model of inflation (say in terms of the scalar field
potential at the brane). Already the simplest model of inflation in the
RSI scenario without bulk scalars predicts that $T/S$ may be different
from that predicted in the four dimensional theory. The theory of
scalar perturbations in braneworld scenarios with bulk scalars is even
more involved. In general, we expect that $T/S$ from braneworld
inflation will be model-dependent and depend on many parameters.

Finally, we will make a comment on the transplanckian problem in the
context of extra dimensions plus inflation. The idea beyond this
problem is the following \cite{trans}: In an expanding universe the
physical momentum of quantum fluctuations is redshifted to smaller
values. Going backwards in time, we have to deal with ever increasing
momenta which at some instant will be equal to the fundamental mass
scale $M_S$. At this point and beyond the quantum field theory approach
to fluctuations in an expanding universe (say during inflation) should
be replaced by a string theory description, which is not yet available.
For quantum fluctuations from inflation in the 4d theory there is,
however, a convenient phenomenological coding of the problem in terms
of the Bogolyubov coefficients $\alpha_k, \beta_k$ for the initial
vacuum state: instead of the vacuum eigenmode $f_k(\tau)$ at infinity
(given in terms of the Hankel function (\ref{temp1})) one can use the
mode function
\begin{equation}\label{mode}
  f_k(\tau) \to \alpha_k f_k(\tau)+\beta_k f_k^*(\tau), \hspace{1em}
  |\alpha_k|^2-|\beta_k|^2,
\end{equation}
and discuss further the effects from $\beta_k$, if any.
If the sizes of the extra dimensions are larger than
$M_S^{-1}$ which takes place in many phenomenological schemes (like the
HW theory), then we must simultaneously deal with extra dimensions and
the transplanckian problem. It is, however, straightforward to combine
the formalism of cosmological fluctuations and that of the
transplanckian problem. If the sizes of the extra dimensions are large
enough to work there in the supergravity approximation, we can use the
quantum field theory of fluctuations in the bulk, as we did in
this paper. Then the transplanckian story will again be coded in terms
of $\beta_k$ for bulk quantum eigenmodes (of gravitons, for instance).

\section*{Acknowledgments}

We are grateful to R. Brandenberger, G. Felder, Y. Dolivet, W. Israel, A. Linde,
S. Mukohyama, S. Pichler, A. Starobinsky and D. Wands for valuable discussions. This
research was supported in part by the Natural Sciences and Engineering
Research Council of Canada and CIAR.

%%% Bibliography


\begin{references}

\bibitem{starob}
%\bibitem{Starobinsky:ty}
A.~A.~Starobinsky,
{\it Spectrum of relict gravitational radiation and the early state of the universe},
JETP Lett.\  {\bf 30}, 682 (1979)
[Pisma Zh.\ Eksp.\ Teor.\ Fiz.\  {\bf 30}, 719 (1979)].
%%CITATION = JTPLA,30,682;%%

\bibitem{drum}
M.~Kac,
{\it Can one hear the shape of a drum?},
Amer.\ Math.\ Monthly, 73 (1966).

\bibitem{brane}
%\bibitem{Horava:1995qa}
P.~Horava and E.~Witten,
{\it Heterotic and type I string dynamics from eleven dimensions},
Nucl.\ Phys.\ B {\bf 460}, 506 (1996)
[{\tt hep-th/9510209}];
%%CITATION = HEP-TH 9510209;%%
%\bibitem{Lukas:1998tt}
A.~Lukas, B.~A.~Ovrut, K.~S.~Stelle and D.~Waldram,
{\it Heterotic M-theory in five dimensions},
Nucl.\ Phys.\ B {\bf 552}, 246 (1999)
[{\tt hep-th/9806051}];
%%CITATION = HEP-TH 9806051;%%
%\bibitem{Randall:1999ee}
L.~Randall and R.~Sundrum,
{\it A large mass hierarchy from a small extra dimension},
Phys.\ Rev.\ Lett.\  {\bf 83}, 3370 (1999)
[{\tt hep-ph/9905221}];
%%CITATION = HEP-PH 9905221;%%
%\bibitem{Randall:1999vf}
L.~Randall and R.~Sundrum,
{\it An alternative to compactification},
Phys.\ Rev.\ Lett.\  {\bf 83}, 4690 (1999)
[{\tt hep-th/9906064}];
%%CITATION = HEP-TH 9906064;%%
%\bibitem{Goldberger:1999uk}
W.~D.~Goldberger and M.~B.~Wise,
{\it Modulus stabilization with bulk fields},
Phys.\ Rev.\ Lett.\  {\bf 83}, 4922 (1999)
[{\tt hep-ph/9907447}];
%%CITATION = HEP-PH 9907447;%%
%\bibitem{DeWolfe:1999cp}
O.~DeWolfe, D.~Z.~Freedman, S.~S.~Gubser and A.~Karch,
{\it Modeling the fifth dimension with scalars and gravity},
Phys.\ Rev.\ D {\bf 62}, 046008 (2000)
[{\tt hep-th/9909134}].
%%CITATION = HEP-TH 9909134;%%

\bibitem{form}
%\bibitem{Mukohyama:2000ui}
S.~Mukohyama,
{\it Gauge-invariant gravitational perturbations of maximally symmetric spacetimes},
Phys.\ Rev.\ D {\bf 62}, 084015 (2000)
[{\tt hep-th/0004067}];
%%CITATION = HEP-TH 0004067;%%
%\bibitem{Kodama:2000fa}
H.~Kodama, A.~Ishibashi and O.~Seto,
{\it Brane world cosmology: Gauge-invariant formalism for perturbation},
Phys.\ Rev.\ D {\bf 62}, 064022 (2000)
[{\tt hep-th/0004160}];
%%CITATION = HEP-TH 0004160;%%
%\bibitem{Langlois:2000ia}
D.~Langlois,
{\it Brane cosmological perturbations},
Phys.\ Rev.\ D {\bf 62}, 126012 (2000)
[{\tt hep-th/0005025}];
%%CITATION = HEP-TH 0005025;%%
%\bibitem{vandeBruck:2000ju}
C.~van de Bruck, M.~Dorca, R.~H.~Brandenberger and A.~Lukas,
{\it Cosmological perturbations in brane-world theories: Formalism},
Phys.\ Rev.\ D {\bf 62}, 123515 (2000)
[{\tt hep-th/0005032}];
%%CITATION = HEP-TH 0005032;%%
%\bibitem{Koyama:2000cc}
K.~Koyama and J.~Soda,
{\it Evolution of cosmological perturbations in the brane world},
Phys.\ Rev.\ D {\bf 62}, 123502 (2000)
[{\tt hep-th/0005239}];
%%CITATION = HEP-TH 0005239;%%
%\bibitem{Deruelle:2000yj}
N.~Deruelle, T.~Dolezel and J.~Katz,
{\it Perturbations of brane worlds},
Phys.\ Rev.\ D {\bf 63}, 083513 (2001)
[{\tt hep-th/0010215}].
%%CITATION = HEP-TH 0010215;%%

\bibitem{HHR}
%\bibitem{Hawking:2000kj}
S.~W.~Hawking, T.~Hertog and H.~S.~Reall,
{\it Brane new world},
Phys.\ Rev.\ D {\bf 62}, 043501 (2000)
[{\tt hep-th/0003052}].
%%CITATION = HEP-TH 0003052;%%

\bibitem{LMW}
%\bibitem{Langlois:2000ns}
D.~Langlois, R.~Maartens and D.~Wands,
{\it Gravitational waves from inflation on the brane},
Phys.\ Lett.\ B {\bf 489}, 259 (2000)
[{\tt hep-th/0006007}].
%%CITATION = HEP-TH 0006007;%%

\bibitem{GKL}
%\bibitem{Giudice:2002vh}
G.~F.~Giudice, E.~W.~Kolb, J.~Lesgourgues and A.~Riotto,
{\it Transdimensional physics and inflation},
[{\tt hep-ph/0207145}].
%%CITATION = HEP-PH 0207145;%%

\bibitem{KS84}
%\bibitem{Kodama:bj}
H.~Kodama and M.~Sasaki,
{\it Cosmological perturbation theory},
Prog.\ Theor.\ Phys.\ Suppl.\  {\bf 78}, 1 (1984).
%%CITATION = PTPSA,78,1;%%


\bibitem{higuchi}
%\bibitem{Higuchi:1986wu}
A.~Higuchi,
{\it Symmetric tensor spherical harmonics on the $N$-sphere and their application to the de Sitter group $SO(N,1)$},
J.\ Math.\ Phys.\  {\bf 28}, 1553 (1987);
%%CITATION = JMAPA,28,1553;%%
%\bibitem{Higuchi:tk}
A.~Higuchi,
{\it Quantum linearization instabilities of de Sitter space-time. 1},
Class.\ Quant.\ Grav.\  {\bf 8}, 1961 (1991);
%%CITATION = CQGRD,8,1961;%%
%\bibitem{Higuchi:tm}
A.~Higuchi,
{\it Quantum linearization instabilities of de Sitter space-time. 2},
Class.\ Quant.\ Grav.\  {\bf 8}, 1983 (1991);
%%CITATION = CQGRD,8,1983;%%
%\bibitem{Higuchi:tn}
A.~Higuchi,
{\it Linearized gravity in de Sitter space-time as a representation of $SO(4,1)$},
Class.\ Quant.\ Grav.\  {\bf 8}, 2005 (1991).
%%CITATION = CQGRD,8,2005;%%

\bibitem{FFK}
%\bibitem{Felder:2001da}
G.~N.~Felder, A.~Frolov and L.~Kofman,
{\it Warped geometry of brane worlds},
Class.\ Quant.\ Grav.\  {\bf 19}, 2983 (2002)
[{\tt hep-th/0112165}].
%%CITATION = HEP-TH 0112165;%%

\bibitem{allen}
%\bibitem{Allen:1987bk}
B.~Allen,
{\it The stochastic gravity wave background in inflationary universe models},
Phys.\ Rev.\ D {\bf 37}, 2078 (1988).
%%CITATION = PHRVA,D37,2078;%%

\bibitem{KS87}
%\bibitem{Kofman:ie}
L.~A.~Kofman and A.~A.~Starobinsky,
{\it Instability of a scalar field in a geometry with anisotropic inflation},
Phys.\ Lett.\ B {\bf 188}, 399 (1987).
%%CITATION = PHLTA,B188,399;%%

\bibitem{KL}
%\bibitem{Kaloper:1998sw}
N.~Kaloper and A.~D.~Linde,
{\it Inflation and large internal dimensions},
Phys.\ Rev.\ D {\bf 59}, 101303 (1999)
[{\tt hep-th/9811141}].
%%CITATION = HEP-TH 9811141;%%

\bibitem{trans}
G.~Chibisov and Yu.~Shtanov,
{\it Structural anisotropy in a chaotic inflationary universe},
Sov.\ Phys.\ JETP {\bf 69}, 17 (1989);
%\bibitem{Brandenberger:2000wr}
R.~H.~Brandenberger and J.~Martin,
{\it The robustness of inflation to changes in super-Planck-scale physics},
Mod.\ Phys.\ Lett.\ A {\bf 16}, 999 (2001)
[{\tt astro-ph/0005432}];
%%CITATION = ASTRO-PH 0005432;%%
%\bibitem{Starobinsky:2001kn}
A.~A.~Starobinsky,
{\it Robustness of the inflationary perturbation spectrum to trans-Planckian  physics},
Pisma Zh.\ Eksp.\ Teor.\ Fiz.\  {\bf 73}, 415 (2001)
[JETP Lett.\  {\bf 73}, 371 (2001)]
[{\tt astro-ph/0104043}];
%%CITATION = ASTRO-PH 0104043;%%
%\bibitem{Kaloper:2002uj}
N.~Kaloper, M.~Kleban, A.~E.~Lawrence and S.~Shenker,
{\it Signatures of short distance physics in the cosmic microwave background},
[{\tt hep-th/0201158}];
%%CITATION = HEP-TH 0201158;%%
%\bibitem{Brandenberger:2002hs}
R.~H.~Brandenberger and J.~Martin,
{\it On signatures of short distance physics in the cosmic microwave background},
[{\tt hep-th/0202142}];
%%CITATION = HEP-TH 0202142;%%
%\bibitem{Danielsson:2002kx}
%U.~H.~Danielsson,
%{\it A note on inflation and transplanckian physics},
%Phys.\ Rev.\ D {\bf 66}, 023511 (2002)
%[{\tt hep-th/0203198}];
%%CITATION = HEP-TH 0203198;%%
%\bibitem{Easther:2002xe}
R.~Easther, B.~R.~Greene, W.~H.~Kinney and G.~Shiu,
{\it A generic estimate of trans-Planckian modifications to the primordial power spectrum in inflation},
Phys.\ Rev.\ D {\bf 66}, 023518 (2002)
[{\tt hep-th/0204129}].
%%CITATION = HEP-TH 0204129;%%
%\bibitem{Starobinsky:2002rp}
%A.~A.~Starobinsky and I.~I.~Tkachev,
%{\it Trans-Planckian particle creation in cosmology and ultra-high energy cosmic rays},
%[{\tt astro-ph/0207572}].
%%CITATION = ASTRO-PH 0207572;%%

\end{references}
\end{document}